\documentclass[aps,prl,twocolumn,groupedaddress]{revtex4}

\usepackage{dcolumn}
\usepackage{bm}
\usepackage{graphicx}
\usepackage{epstopdf}
\usepackage{array}
\usepackage{xcolor}
\newcolumntype{L}[1]{>{\raggedright\let\newline\\\arraybackslash\hspace{0pt}}m{#1}}
\newcolumntype{C}[1]{>{\centering\let\newline\\\arraybackslash\hspace{0pt}}m{#1}}
\newcolumntype{R}[1]{>{\raggedleft\let\newline\\\arraybackslash\hspace{0pt}}m{#1}}
\begin{document}

\title{Synaptic-Like Plasticity in 2D Nanofluidic Memristor from Competitive Bicationic Transport}

\author{Yechan Noh}
\affiliation{Department of Physics, University of Colorado Boulder, Boulder, CO 80309, USA\footnote{Current affiliation: University of Colorado Boulder and National Institute of Standards and Technology}}
\affiliation{Applied Chemicals and Materials Division, National Institute of Standards and Technology, Boulder, CO 80305, USA$^*$}
\affiliation{Department of Materials Science and Engineering, University of California Berkeley, Berkeley, CA 94720, USA}
\author{Alex Smolyanitsky}
\email[Corresponding author: ]{alex.smolyanitsky@nist.gov}
\affiliation{Applied Chemicals and Materials Division, National Institute of Standards and Technology, Boulder, CO 80305, USA}

\date{\today}

\begin{abstract} Synaptic plasticity, the dynamic tuning of signal transmission strength between neurons, serves as a fundamental basis for memory and learning in biological organisms. This adaptive nature of synapses is considered one of the key features contributing to the superior energy efficiency of the brain. In this study, we utilize molecular dynamics simulations to demonstrate synaptic-like plasticity in a subnanoporous 2D membrane. We show that a train of voltage spikes dynamically modifies the membrane's ionic permeability in a process involving competitive bicationic transport. This process is shown to be repeatable after a given resting period. Due to a combination of sub-nm pore size and the atomic thinness of the membrane, this system exhibits energy dissipation of 0.1--100 aJ per voltage spike, which is several orders of magnitude lower than 0.1--10 fJ per spike in the human synapse. We reveal the underlying physical mechanisms at molecular detail and investigate the local energetics underlying this apparent synaptic-like behavior.
\end{abstract}

\maketitle
Synapses are junctions that enable chemo-electrical signaling between neurons. In a typical synapse, the signal transmission strength is dynamically modulated in response to previous neural activity, a feature referred to as synaptic plasticity~\cite{abbott2000synaptic, martin2000synaptic}. This adaptive alteration of synaptic strength plays a fundamental role in memory and learning functions in living organisms. Moreover, it enables biological neural networks to concurrently perform both processing and storage of information in a sparse manner, a feature believed to be central to their superior energy efficiency. Inspired by these biological functionalities, artificial electrical elements with synaptic-like plasticity have been studied extensively~\cite{huh2020memristors, markovic2020physics, sebastian2020memory, noy2023fluid, chen2023electrochemical, lee2021organic}, aimed at building analog artificial neural networks with substantially enhanced energy efficiency compared to emulations based on the von Neumann computing architecture.

Memristors~\cite{1083337} comprise an area of extensive research due to their potential promise as artificial synaptic elements for neuromorphic computing. Physically, a memristor is an electrical conductor capable of modulating its conductivity in response to previous voltage inputs and maintaining the modulated state without a continuous source of power. This internal gating enables memristor networks to perform information processing and storage simultaneously. Over the past decade, solid-state neuromorphic chips featuring memristor networks have been demonstrated to perform analog machine learning tasks at a fraction of the energy cost of their von Neumann counterparts~\cite{duan2024memristor, sebastian2020memory, ning2023memory, le202364}. More recently, there has been a spike of interest in nanofluidic memristors to directly mimic biological neural networks~\cite{bu2019nanofluidic, sheng2017transporting, zhang2019nanochannel, robin2021modeling, xiong2023neuromorphic, robin2023long, emmerich2024nanofluidic, paulo2023hydrophobically, shi2023ultralow, noh2024memristive, zhou2024nanofluidic, ramirez2024memristive, ramirez2024neuromorphic}, with two recent works notably reporting long-term memory effects along with basic Hebbian learning~\cite{robin2023long}), as well as operating voltage comparable to that of biological synapses and yielding sub-picojoule energy consumption per spike ~\cite{xiong2023neuromorphic}. In nanofluidic memristors, aqueous ions serve as the charge carriers instead of electrons, in resemblance to the human brain. One noteworthy difference between electrons and ions as charge carriers is the rich diversity of the latter, which can coexist within a given system. In particular, the competitive interplay between the ionic species leads to interesting phenomena, including ion sieving~\cite{fang2019highly} and memristive ion transport~\cite{noh2024memristive}. It is well known that biological systems readily harness the diversity of ion species for their functions, as most notably exemplified by the generation of action potentials in neurons~\cite{bean2007action}. Therefore, exploring ways to harness ionic diversity in artificial nanofluidic systems to achieve neuromorphic functions may represent a crucial research direction in nanofluidics. 

Among fluidic ion conductors, nanoporous 2D membranes represent a class of materials with high energy efficiency of ion transport. The primary reason for this efficiency is the atomic thinness, which, combined with sub-nm pore dimensions, enables relatively low, highly localized and ion-selective permeation barriers, as described earlier ~\cite{Smolyanitsky2018,fang2019highly,fang2019mos2,hassani2021gas}. Unsurprisingly, this class of materials has been considered for a range of applications, including water desalination~\cite{liu2016two, cohen2012water, yang2019large, noh2021phonon}, molecular separation~\cite{liu2016two, zhu2018rapid, xu2022highly}, and osmotic energy harvesting~\cite{feng2016single, safaei2022progress}. Combined with high permeation selectivity and the prospect of high-density pore array fabrication~\cite{moreno2018bottom, choi2018multifunctional, zhong2019wafer}, subnanoporous 2D membranes appear to be excellent candidates for artificial synaptic devices. Achieving reliable memory functionality, however, is neither trivial, nor necessarily intuitive. In this study, we demonstrate synaptic-like plasticity of aqueous ion transport through a sub-nanoporous 2D membrane. We demonstrate that it arises from the adsorption/desorption and transport of two cation species with markedly different ion-pore affinities. Finally, We provide a comprehensive molecular-level insight into the underlying mechanisms.

\begin{figure*}[ht]
\centering
\includegraphics[width=0.8\textwidth]{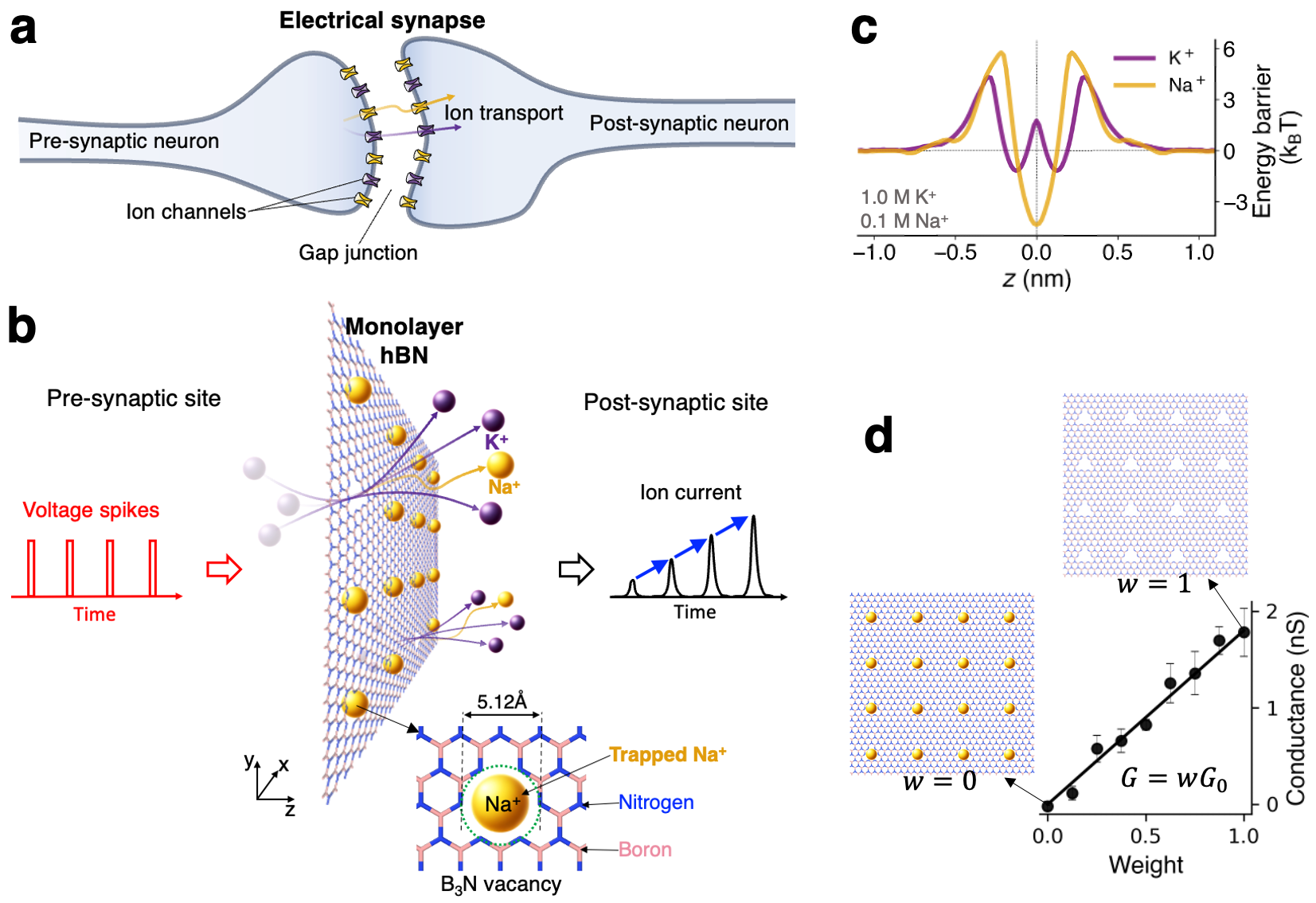}
\caption{ \textbf{Sketch of the system and characteristics of ion transport across a subnanoporous hBN monolayer} \textbf{a.} Illustration of biological electrical synapse and \textbf{b.} 2D porous membrane involving bicationic ion transport driven by spiking voltage, which causes desorption of Na$^{+}$ ions from the pores and activating K$^{+}$ transport. \textbf{c.} Energy profile for K$^{+}$ and Na$^{+}$ ions along the transport coordinate for a binary salt mixture containing 1.0 M KCl and 0.1 M NaCl. \textbf{d.} Membrane conductance \textit{vs} weight at 1.0 M K$^{+}$. The insets in (d) show Na$^{+}$ trapping states at the corresponding weight values.}
\label{fig1}
\vspace{12pt} 
\end{figure*}

We used all-atom molecular dynamics (MD) simulations to investigate dynamic ion transport across a 2D porous membrane under a sequence of rectangular voltage pulses. Fig.~\ref{fig1}a shows a sketch of ion transport in a biological electrical synapse, where ions are transported between the ion channels across the gap junction between the pre-synaptic neuron and the post-synaptic neuron. The plasticity of such synapses typically arises from the dynamic changes in conductance of the voltage-gated channels, as well as the gap region~\cite{curti2016characteristics, harris2001emerging}. In this study, we considered a sub-nanoporous 2D membrane mimicking a simplified artificial synapse. Fig.~\ref{fig1}b illustrates bicationic (Na$^{+}$ and K$^{+}$) ion transport through an array of sub-nanometer pores hosted by a 2D membrane. In this case, the membrane material is a hexagonal boron nitride (hBN) monolayer with a total of 16 regularly spaced B${_3}$N triangular multivacancy pores within an area of approximately 7 nm $\times$ 7 nm. The B${_3}$N pore is a defect commonly found in monolayer hBN~\cite{park2021atomically} and its effects on wettability~\cite{kumar2022surface}, water slippage~\cite{seal2021modulating}, as well as ion trapping and mechanosensitive ion transport~\cite{noh2024Stretch-Inactivated} have been studied. The membrane is suspended in the middle of a simulation box and immersed in a binary mixture of water-dissociated 1.0 M KCl and 0.1 M NaCl, unless the concentrations are stated otherwise. Dynamic transport response in this system is initiated by a sequence of rectangular electric field pulses externally applied in the $Z$-direction, as shown in Fig.~\ref{fig1}b. Further details on the simulation procedures can be found in the Methods section.

The permeation properties of B${_3}$N vacancies in hBN are worth introducing first. These electrically neutral sub-nm pores feature dipolar electrostatics with negatively charged dipole components located at the edge nitrogen atoms, enabling selective transport. More specifically, anions are outright rejected, while K$^{+}$ ions permeate relatively rapidly and Na$^{+}$ ions hardly permeate as they become stably trapped in the pores~\cite{noh2024Stretch-Inactivated}. Fig.~\ref{fig1}c shows the corresponding free energy landscapes in the form of potentials of mean force (PMF) along the transport coordinate ($Z$) for both ion species. For Na$^{+}$, the energy profile features a barrier of $\approx10k_BT$ at $Z=0$ ($k_B$ is the Boltzmann constant and $T=300$ K is the system temperature). The K$^{+}$ ions, however, experience a much shallower energy well, $\approx5.5k_BT$-deep, resulting in relatively weak trapping of K$^{+}$ ions. As described earlier~\cite{Smolyanitsky2018,fang2019highly,fang2019mos2}, the rate of ion transport through these barrier-limited pores is of the Arrhenius type, \textit{i.e.}, $I \propto e^{-\frac{\Delta E}{k_BT}}$, where $\Delta E$ is the rate-setting peak-to-peak free energy barrier. Unsurprisingly, given an energy barrier difference of $\approx4.5k_BT$, the rate of K$^{+}$ permeation is nearly two orders of magnitude higher than that of Na$^{+}$ in a single-salt scenario. Following from the same argument, Na$^{+}$ ions spend significantly more time trapped inside the pores, compared to K$^{+}$, which overall is similar to 18-crown-6 ether pores in graphene, except the ion-pore affinity is reversed for the same cation pair~\cite{Smolyanitsky2018,fang2019highly}. Given that a pore of this size becomes impermeable when a cation is trapped inside, the number of available conductive paths is the total number of pores \textit{unoccupied} by Na$^+$. The effective membrane conductance for a salt mixture thus can be written as a simple linearly weighted function proposed earlier~\cite{noh2024memristive}, directly evaluated in Fig.~\ref{fig1}d:
\begin{equation}
G = wG_0,
\label{eq1}
\end{equation}
where $G_0$ is the conductance of a membrane completely deoccupied by Na$^+$. The corresponding weight $w$ is the time-dependent fraction of empty pores: $w(t) = 1 - {N(t)}/{N_{tot}}$, where $N(t)$ is the number of pores ``plugged'' by Na$^+$ and $N_{tot}$ is the total number of pores in the array.

\begin{figure*}[ht]
\centering
\includegraphics[width=1.0\textwidth]{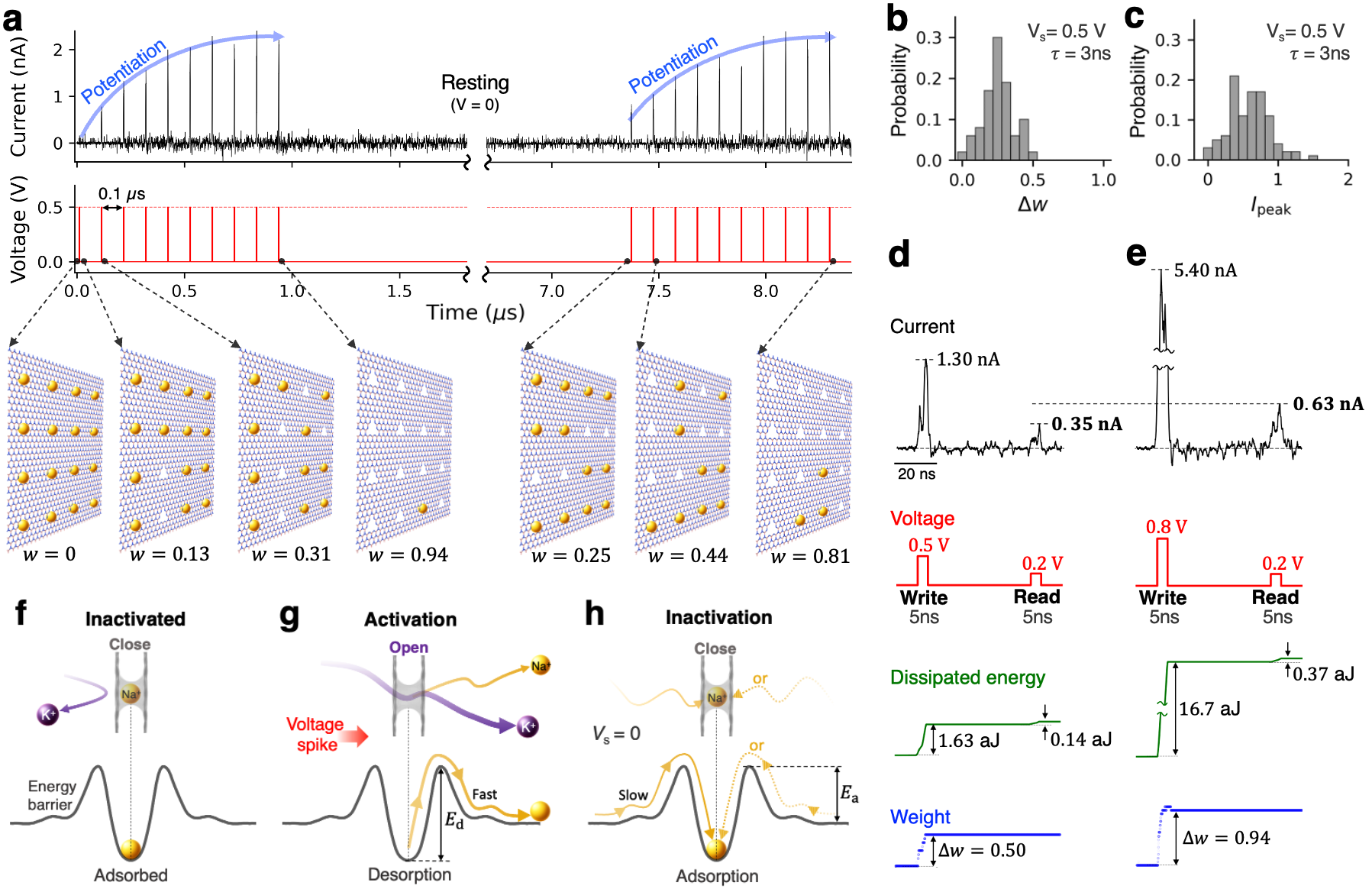}
\caption{  \textbf{Synaptic-like plasticity in ion transport across a hBN monolayer membrane.} \textbf{a.} Ion current potentiation by rectangular voltage pulses. In the resting cycle, the system is unbiased. The subfigures illustrate trapped Na$^{+}$ ions at selected times. The probability distribution of \textbf{b.} pulse-induced weight increments and \textbf{c.} current peaks obtained from 100 independent simulations with a 0.5 V voltage pulse of $\tau$ = 3 ns. Writing and reading operations using a write pulse of \textbf{d.} 0.5 V and \textbf{e.} 0.8 V, followed by a 0.2 V read pulse. All read and writepulses were 5 ns long. The interval between writing and reading operations is 50 ns. Mechanisms associated with the synaptic-like plasticity: \textbf{f.} Inactivated state: the trapped Na$^{+}$ blocks K$^{+}$ transport. \textbf{g.} Activation or learning process: bias-induced desorption of Na$^{+}$. \textbf{h.} Inactivation or forgetting process: readsorption of Na$^{+}$ ions}
\label{fig2}
\vspace{12pt} 
\end{figure*}

Shown in Fig.~\ref{fig2} are the results for synaptic-like plasticity exhibited by the aqueous ion transport through an array of B$_3$N vacancies in an 2D hBN monolayer under a pulsed bias. In the context of this work, plasticity refers to the dynamic alteration of the membrane's ionic permeability in response to voltage bias history. We first applied ten 3-ns-long, 0.5V voltage pulses with an interval of 0.1 $\mu$s (results obtained for different intervals can be found in the supplementary Fig. 1). Before the first voltage pulse, none of the pores were activated \textit{i.e.,} all 16 pores were occupied by the trapped Na$^{+}$ ions. The pore activation dynamics as a function of subsequent voltage pulses is shown in Fig.~\ref{fig2}a, with the first pulse activating two pores and second pulse activating three more pores, corresponding to $w=2/16=0.125$ and $w=5/16=0.3125$, respectively--and so forth. At a finite temperature, the individual $\Delta w$ increments are of course stochastic and thus the analytical discussions provided later in the text correspond to the statistically significant $\Delta w$ values, \textit{i.e.}, those obtainable from \textit{repeated} identical pulse trains applied to a system initially at $w=0$. The distributions of weight change and peak current ($I_{peak}$) resulting from a 3-ns-long 0.5V pulse are shown in Figs.~\ref{fig2}b and c, respectively. The corresponding average weight change is 0.256 with a standard deviation of 0.108 for a membrane featuring 16 pores. This type of stochasticity introduces a degree of inherent randomness, which results in natural diversity in conductance switching. In principle, intrinsic stochasticity is commonly found in solid-state memristors~\cite{duan2024memristor} and the human brain~\cite{markovic2020physics}. Specifically for the presented system, the degree of randomness should decrease with increasing array size. As successive voltage pulses are applied, these spikes expectedly exhibit an increasing trend with a clear asymptote corresponding to the maximum weight $w=1$ (at the end of the first 10-pulse sequence, the value of $w$ increased to 0.94, corresponding to only one trapped Na$^+$ ion). The information about the prior history of voltage pulses is stored in the form of cumulatively added deoccupancy of Na$^{+}$. This gradual potentiation of ion permeability arises due to the competitive transport between K$^{+}$ and Na$^{+}$, and thus the potentiation effects do not take place with a monocationic electrolyte (see Supplementary Fig. 2). As expected for a system symmetric with respect to the membrame, the potentiation occurs regardless of bias reversal, resulting in bidirectional memristive behavior described in Supplementary Fig. 3. Once the pulse train stops, the system undergoes resting, during which the Na$^{+}$ ions are slowly re-trapped, volatilizing the previously gained memory. After approximately 6 $\mu$s of rest, 11 additional Na$^{+}$ ions are re-trapped, leaving only four pores empty and corresponding to $w=4/16=0.25$. After that, the system underwent another learning cycle with the same sequence of voltage pulses, exhibiting ion current potentiation similar to that observed in the first cycle, except with a different initial value of $w$. 
As demonstrated, the conductance state of the device is switchable by voltage pulses a few nanoseconds in duration, resulting in a switching resolution at the scale of order 0.1 GHz. The relaxation time of a few microseconds, on the other hand, corresponds to the effective memory retention time, tunable by the local association barriers, as well as the ion concentrations, as discussed in greater detail further in the text. These timing ranges may therefore be of specific interest for implementing devices that aim to combine GHz-scale state switching with MHz-scale state retention.

Given the discussion above, the high-magnitude component of the bias pulse can be viewed as the write operation performed on a membrane, while a measurement of the ion current during the lower-magnitude pulse is the read operation aimed at probing the membrane's ionic permeability. For writing, we used 5-ns-long pulses of relatively high magnitude (0.5 V and higher), which enable rapid removal of some of the Na$^+$ ions from the pores. For reading, pulses of lower magnitude (0.2 V) were used, allowing to merely probe the membrane permeability in its Na$^+$-occupancy state without modifying the latter. Figs.~\ref{fig2}d and ~\ref{fig2}e show the corresponding results for write-pulses of 0.5 V and 0.8 V, followed by a read-pulse after 50 ns. As expected, permeability potentiation is stronger with a write-pulse of higher magnitude: $\Delta w=0.50$ and $\Delta w=0.94$ for 0.5 V and 0.8 V pulse, respectively. The corresponding dissipated energy expenditure is quite low. We define the energy dissipated in a given pulse as $\Delta E= \int_{0}^{\tau} I(t) V_s dt$, where $I(t)$ is the ion current and $V_s$ is the pulse height, integrated for the pulse duration $\tau$. As shown in Figs.~\ref{fig2}g and ~\ref{fig2}g, the  energy dissipation per voltage spike in this 2D nanofluidic memristor is on the attojoule scale, attributed to the relatively low ion currents and the nanosecond-scale pulse. These energy estimates are several orders of magnitude lower than those of human synapses, which have an energy expenditure of roughly 0.1-10 fJ per synaptic event~\cite{lee2021organic, noy2023fluid, laughlin1998metabolic}. Note that the attojoule-scale dissipation estimate given above should only be viewed as an idealized lower bound. In a more realistic scenario, especially given the current state-of-the art fabrication techniques, parasitic effects in the form of ion current leakage and other sources would certainly increase dissipative losses. At the same time, this lower bound may serve as a suitable objective in terms of device design and fabrication.

The main mechanism underlying the simple potentiating-forgetting cycle described above is dynamic adsorption/desorption of Na$^+$ ions by the pores, with inactivated and activated states of pores as sketched in Figs.~\ref{fig2}f and ~\ref{fig2}g, respectively. When a write-pulse is applied, a significant probability of Na$^+$ desorption arises, as set by the write-pulse voltage peak and the desorption barrier $E_d$ (see Fig.~\ref{fig2}g for the definition). Upon forgetting, adsorption occurs with a considerably lower probability, as set by the adsorption barrier $E_a$ (see Fig.~\ref{fig2}g) and the low concentration of Na$^+$ ions. In particular, to achieve low memory volatility (\textit{i.e.}, in the form of a long forgetting time in Fig.~\ref{fig2}a), a combination of relatively high $E_a$ and sufficiently low Na$^+$ concentration is essential, because the corresponding adsorption rate that describes the spontaneous ``forgetting'' process is $\propto c_{Na^{+}}\exp(-{\frac{E_a}{k_B T})}$, where $c_{Na^{+}}$ is the sodium concentration.  

\begin{figure*}[ht]
\centering
\includegraphics[width=0.95\textwidth]{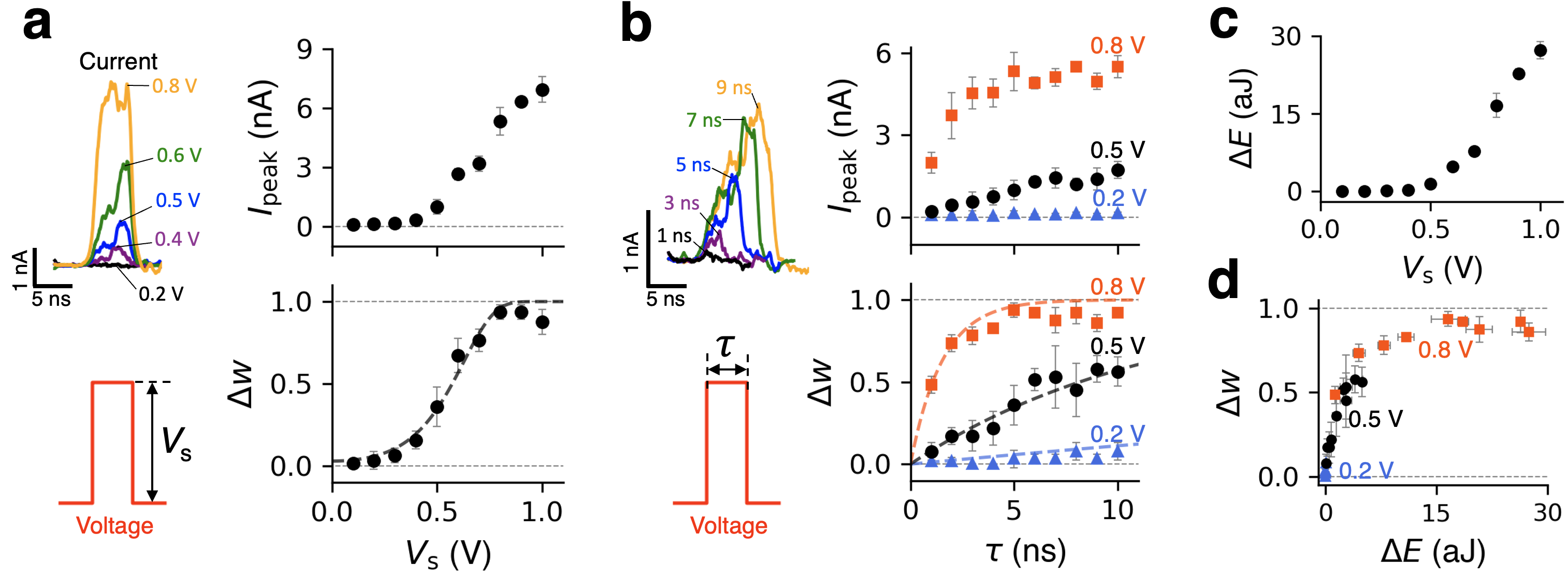}
\caption{\textbf{Synaptic potentiation and energy consumption during a single voltage pulse of varying magnitude and duration.} Peak current and weight change for \textbf{a.} varying $V_s$ with at fixed $\tau$ = 5 ns and \textbf{b.} varying $\tau$ at fixed $V_s$. \textbf{c.} Dissipated energy as a function of $V_s$ with a fixed $\tau$ = 5 ns \textbf{d.} Dissipated energy \textit{vs} weight change. As shown, $I_{peak}$, $\Delta$w, and $\Delta$E are averages from four independent simulations. The error bars are the corresponding standard deviations.}
\label{fig3}
\vspace{12pt} 
\end{figure*}

A simple analytical model describing the entire process presented above is possible. For an individual pore, the effective desorption rate in the presence of external bias is given by $r_d = f_d \cosh\left({\frac{q\phi}{2k_B T}}\right) \exp\left({-\frac{E_d}{k_B T}}\right)$, where $f_d$ is the attempt frequency associated with thermal fluctuations, $\phi \propto V(t)$ is the bias-induced local shift of the ion's electrostatic potential, and $q$ is the electric charge of the ion. As mentioned earlier, the concentration-dependent adsorption rate is $r_a = \kappa c_{Na^{+}} \exp\left({-\frac{E_a}{k_B T}}\right)$ (see Fig.~\ref{fig2}g for the definition of $E_a$), where $c_{Na^{+}}$ is the concentration of Na$^+$ ions and $\kappa$ is a suitable transmission coefficient, such that $\kappa c_{Na^{+}}$ is the corresponding adsorption attempt frequency. In general, $r_a$ is bias-dependent, but we posit that the dependence here is considerably weaker than for desorption. 
For a pore array, let us consider the system state determined by the number of Na$^+$ ions trapped in the pores as a function of time, in response to a voltage pulse. We note that despite the fact that the \textit{ability to observe} time delays in the form of ion currents requires at least two cation species with markedly different ion-pore affinities~\cite{noh2024memristive}, the time-delayed state dynamics itself can be described in a single-cation scenario. The number of trapped Na$^+$ ions $N(t)$ satisfies a simple differential equation: $\frac{dN}{dt} = -r_d N + r_a (N_{tot} - N)$. For a constant bias (\textit{e.g.}, during a rectangular voltage pulse), the analytical solution is then $N(t) = N_0 e^{-(r_a+r_d)t} + \frac{N_{tot}r_a}{r_a+r_d}(1 - e^{-(r_a+r_d)t})$, where $N_0$ is the initial state. The corresponding weight is then given by $w = w_0 + \left(\frac{r_d}{r_a+r_d}-w_0\right)(1-e^{-(r_a+r_d)t})$, where $w_0 = 1-N_0/N_{tot}$ is the initial weight. Given that $r_d$ is an exponential function of the external bias, two distinct processes are possible, depending on the relative strengths of $r_a$ and $r_d$. 

As shown earlier, the system ``learns'' from a train of ``writing'' pulses when $r_d\gg r_a$ during each pulse and the interval between the pulses is insufficiently long for significant memory loss to occur. A necessary requirement here is that the bias magnitude is sufficiently high for a pulse lasting only a few nanoseconds to remove a considerable number of sodium ions from the pores. The learning process is then cumulative, \textit{i.e.}, the state after the $n$-th pulse remembers the sum of $w$ changes caused by the previous pulses, until all pores are deoccupied by Na$^+$. After $n$-th pulse of duration $\tau$:  
\begin{equation}
w_n = w_{n-1} + \left( \frac{r_d}{r_a+r_d}-w_{n-1}\right)(1-e^{-(r_a+r_d)\tau}),
\label{eq2}
\end{equation}
where $w_{n-1}$ and $w_n$ is the weight before and after the pulse, respectively. At $r_d\gg r_a$, $w_0=0$, and assuming no appreciable re-adsorption between the equation pulses, the difference equation above yields a geometric series $w_n = 1-e^{-(r_a+r_d)\tau n}$, asymptotically convergent to unity, in accord with the results in Fig.~\ref{fig2}a. Of potential interest, for a train of voltage pulses of constant magnitude and duration, the $w_n - w_{n-1}$ increments are statistically numerically unique. In principle, this suggests the possibility of implicitly ``encoding'' information about the number of pulses that preceded a given unsaturated value of $w$ in the case of large pore arrays. The $r_d$'s sensitivity to the bias magnitude is worth considering in greater detail. Shown in Fig.~\ref{fig3}a are the results of simulated potentiation by a single 5-ns-long voltage pulse of height $V_s$, applied to a membrane initially fully inactivated ($w=0$). Both the current spike height and $\Delta w$ are marked by a threshold in the amount of $\approx0.3 $V, consistent with the definition of $r_d$. Sensitive dependence on the pulse duration $\tau$ at a fixed $V_s$ (see 0.5 V and 0.8 V cases in Fig.~\ref{fig3}b) above the threshold is also shown in Fig.~\ref{fig3}b, causing the corresponding current peak and $\Delta w$ to increase rapidly when $\tau$ increases. In addition, there is reasonable agreement between $\Delta w$ given by Eq.~\ref{eq2} and the simulated data (see the dashed curves in Fig.~\ref{fig3}a and ~\ref{fig3}b). Note that the resulting energy expenditure for various values of $V_s$ shown in Fig.~\ref{fig3}c exhibits a rapid increase above the $V_s$ threshold. In particular, Fig.~\ref{fig3}d shows that energy consumption increases nonlinearly with the weight increment, naturally following the corresponding rapid increase in $I_s$ (compare the top panels in Figs.~\ref{fig3}a and ~\ref{fig3}c). 

\begin{figure*}[ht]
\centering
\includegraphics[width=0.8\textwidth]{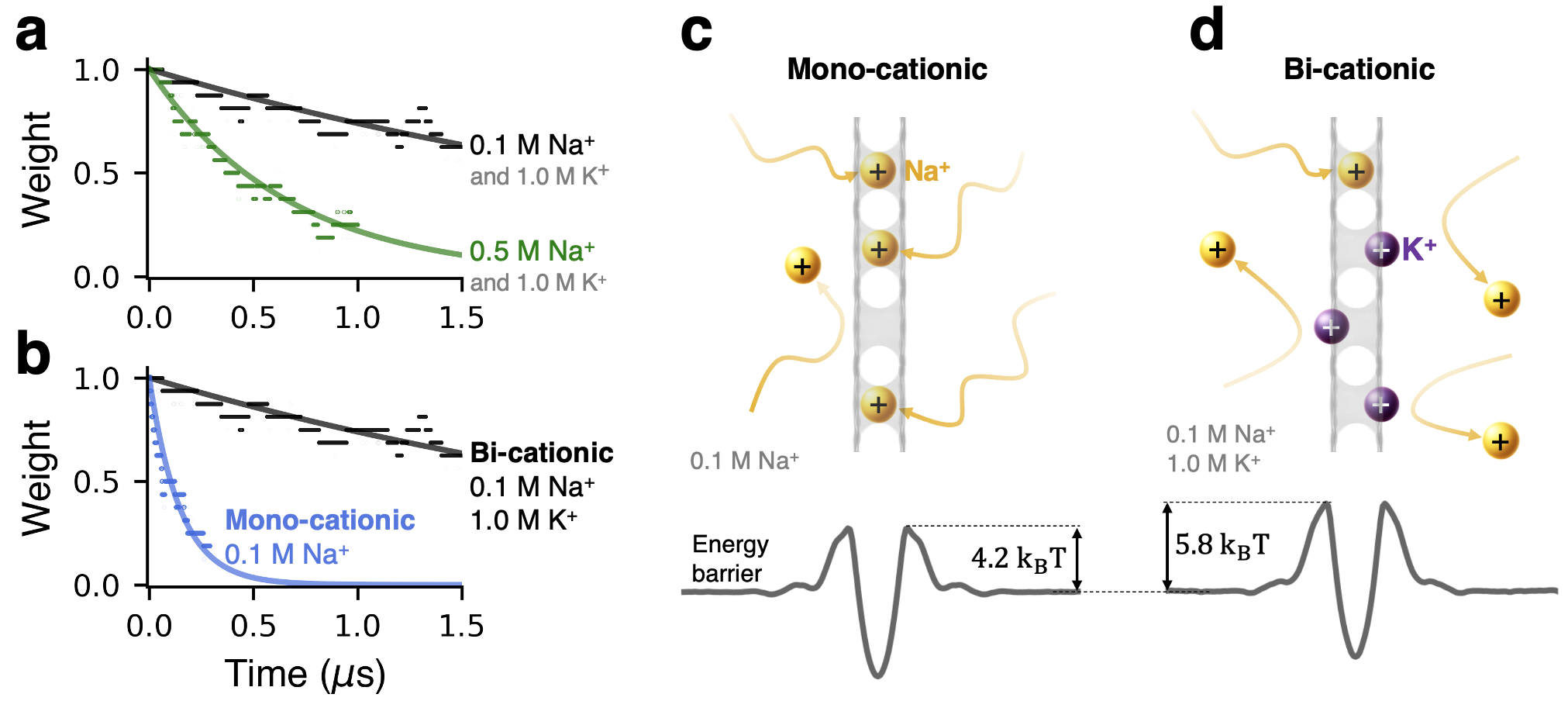}
\caption{\textbf{Effect of Na$^+$ concentration and Na$^+$--K$^+$ interplay in the forgetting process.} \textbf{a.} Effect of Na$^{+}$ concentration on weight decay. \textbf{b.} Effect of K$^{+}$ ions on weight decay. Sketches of the forgetting process for \textbf{c.} mono-cationic electrolyte featuring only Na$^{+}$ and \textbf{d.} bicationic electrolyte containing Na$^{+}$ and K$^{+}$ at the stated ion concentrations, along with the corresponding energy profiles directly below.}
\label{fig4}
\vspace{12pt} 
\end{figure*}

The second important process that can occur is the loss of memory in the longer-term absence of external bias, which manifests as spontaneous gradual decrease in $w$. The unbiased dynamics is described identically to Eq.~\ref{eq2}, except now $r_a$ and $r_d$ are more comparable:
\begin{equation}
w = w_0 + \left(\frac{r_d}{r_a+r_d}-w_0\right)(1-e^{-(r_a+r_d)t}),
\label{eq3}
\end{equation}
where $w_0>0$ is the state prior to the start of memory loss and $t$ is the elapsed time. The decay rate is a sensitive function of $c_{Na^{+}}$ \textit{via} $r_a\propto c_{Na^{+}}$, consistent with the results in Fig.~\ref{fig4}a, which shows spontaneous decay of $w$ at several concentrations of NaCl. The final state of the membrane is also concentration-dependent, because regardless of $w_0$, at $t\rightarrow\infty$ Eq.~\ref{eq3} yields $w = \frac{r_d}{r_a + r_d} = \frac{1}{\lambda c_{Na^+}+1}$, where $\lambda$ is a constant -- also consistent with the results in Fig.~\ref{fig4}a. An important factor affecting the forgetting process omitted in the discussion above is the presence of K$^+$ ions, which are expected to interfere with Na$^+$-pore binding. As shown in Fig.~\ref{fig4}b, the presence of K$^+$ ions indeed slows down the rate of $w$ decay (also see Supplementary section S2). The effects of potassium are in fact deeper than the level of interference captured by our analytical model. Shown in Figs.~\ref{fig4}c and \ref{fig4}d is the effect of K$^+$ on the \textit{barrier heights}, because K$^+$ ions contribute short-range steric and longer-range electrostatic cation-cation repulsion through their presence in the direct vicinity of the pores and near the membrane surfaces, respectively. These observations suggest that the concentration of the main conducting ion species can also be used as a tuning parameter for controlling the system dynamics. 
The possible influence of spurious defects essentially guaranteed to be present in realistic post-fabricaton hBN membranes is worth noting briefly. Smaller defects (\textit{e.g.}, in the form of single-atom or B$_2$N vacancies), are expected to be entirely impermeable to K$^+$ and Na$^+$ ions, because the B$_3$N vacancy is essentially the smallest permeable pore for these cation species. At the same time, smaller cations, such as Li$^+$ and protons, may still be able to permeate through smaller defects. Larger multivacancies, on the other hand, are not expected to trap ions and thus the transport contributions from those pores would likely manifest as baseline leakage current. Although fabrication of large uniform arrays of B$_3$N vacancies indeed remains a challenge, a new fabrication method reported recently shows significant promise in achieving increased control over pore size and shape distribution in an array of multivacancies~\cite{byrne2024probing,byrne2024triangnanopores}.

To summarize, we have demonstrated synaptic-like plasticity of aqueous ion transport through a sub-nanoporous 2D membrane made of a monolayer hBN featuring an array of B$_3$N vacancies in a binary aquous salt. The ion conductance of this nanofluidic memristor can be dynamically modulated by few-nanoseconds-long transmembrane voltage pulses, resulting in a switching speed of approximately 0.1 GHz. The resulting energy dissipation per voltage spike is very low, roughly 0.1—100 aJ per spike, which is several orders of magnitude lower than that of the human brain, 0.1—10 fJ per spike. Notably, this synaptic-like plasticity arises from the dynamic interplay between two cationic species (Na$^{+}$ and K$^{+}$), exemplifying an illustrative approach to leveraging ionic diversity for achieving neuromorphic functionality. The conductive state of the membrane is shown to be governed by the adsorption/desorption dynamics of Na$^{+}$ trapped in the pores, while K$^+$ ions permeate through the empty pores. We derived a set of illustrative analytical expressions associated with the apparent synaptic-like potentiation and memory volatility. Further experimentation is needed to demonstrate the predicted behaviors for various subnanoporous 2D membranes, as well as ionic permeants. It is our hope to stimulate further work aimed at a better understanding of the effects of ionic diversity, including at the system level, \textit{i.e.}, in the case of interconnected nanofluidic memristive elements.

\section*{Methods}
For the MD simulations of ion transport, we used a rectangular box with dimension $L_X=6.998$ nm $\times$ $L_Y=6.926$ nm $\times$ $L_Z=10$ nm, periodic in all directions. The subnanoporous hBN monolayer was placed in the $XY$-plane at $Z=L_Z/2$ with the pore array consisted of 16 triangular B$_3$N vacancies. To prevent the membrane from drifting, its edge atoms were tethered to their initial positions by spring restraints. The system was filled with aqueous salt mixtures consisting of explicitly simulated K$^+$, Na$^+$, and Cl$^-$ ions. As the system was periodic in all directions, the time- and ensemble-averaged salt concentration on each side of the membrane was identical under zero bias. The membrane was simulated using parameters developed earlier~\cite{govind2018ab} within the OPLS-AA forcefield framework~\cite{jorgensen1996development}. The partial charges of the nitrogen atoms at the edges of B$_3$N vacancies were set at $2/3$ of their bulk values (obtained earlier from quantum-chemical calculations~\cite{govind2018ab}), ensuring electrical neutrality of the membrane. Explicitly simulated water molecules were simulated using the TIP4P model~\cite{jorgensen1983comparison}. All of the non-bonded interactions were simulated using the OPLS-AA forcefield framework~\cite{jorgensen1996development}. A particle-particle--particle-mesh scheme was used to simulate the electrostatic interactions. A 1.2 nm cut-off radius was used for all short-range interactions, including short-range Coulomb interactions and the Lennard-Jones interactions. To initialize the conductive weight of the membrane to zero in all of our simulations, we placed Na$^+$ ions inside the B$_3$N vacancies. The system first underwent static energy minimization and then dynamics relaxation in the \textit{NPT} ensemble at $T=300$ K and $P=1$ bar, using a 1 fs time step, with the Parrinello-Rahman barostat modifying simulation box dimensions only in the $Z$-direction. Relaxed systems underwent ion transport simulations under rectangular pulses of external electric field applied in the $Z$-direction, using a 2 fs time step. The corresponding pulse magnitudes were calculated as the electric field magnitude, multiplied by $L_Z$. To obtain the ion currents, we first calculated the cumulative ionic fluxes as a function of time $N(t)$. The flux data was recorded every 10 ps, corresponding to a 100 GHz sampling rate. The raw $N(t)$ data were then filtered using a Chebyshev low-pass digital filter with a 200 MHz cutoff frequency. Finally, time-derivatives of the filtered flux data were obtained using an 8th-order central difference method, yielding the time-dependent ionic current $I(t) = q\frac{dN_{f}(t)}{dt}$, where $N_f(t)$ is the filtered cumulative flux and $q$ is the ionic charge. Further details can be found in the supplementary materials of our previous work~\cite{noh2024memristive}. The PMF data were obtained using the Weighted Histogram Analysis Method~\cite{wham_Hub2010} applied to a total of 60 umbrella samples of the ion's position along the $Z$-coordinate incremented by 0.05 nm; each umbrella sample was obtained from a 20-ns-long simulation. All MD simulations were performed using GPU-accelerated GROMACS~\cite{abraham2015gromacs, pall2020heterogeneous} and the molecular visualization tasks were carried using the OVITO software~\cite{stukowski2009visualization}. 

\section*{Acknowledgments}
A substantial portion of the simulations was conducted using the GPU-accelerated high-performance computing facilities at the National Institute of Standards and Technology. We thank Frances Allen for numerous illuminating discussions regarding fabrication of multivacancy arrays in 2D materials. 
Yechan Noh’s work on subdiffusive transport characterization was supported by National Science Foundation (NSF) under Award No. 2110924. Use of computational resources was supported by NSF under ACCESS program, Award No. NNT230006.

\bibliography{local}

\end{document}


\doublespacing 

\title{Supplementary Material for:

Synaptic-Like Plasticity in 2D Nanofluidic Memristor from Competitive Bicationic Transport}

\author{Yechan Noh}
\affiliation{Department of Physics, University of Colorado Boulder, Boulder, CO 80309, USA}
\affiliation{Applied Chemicals and Materials Division, National Institute of Standards and Technology, Boulder, CO 80305, USA}
\affiliation{Department of Materials Science and Engineering, University of California, Berkeley, Berkeley, CA 94720, USA}
\author{Alex Smolyanitsky}
\affiliation{Applied Chemicals and Materials Division, National Institute of Standards and Technology, Boulder, CO 80305, USA}


\maketitle

\clearpage
\section*{S1. Extended data and discussion of the effects of pulse interval, bicationic electrolytes, bias reversal, bias superposition, and membrane material.}

\begin{figure*}[h]
\centering
\includegraphics[width=1.0\textwidth]{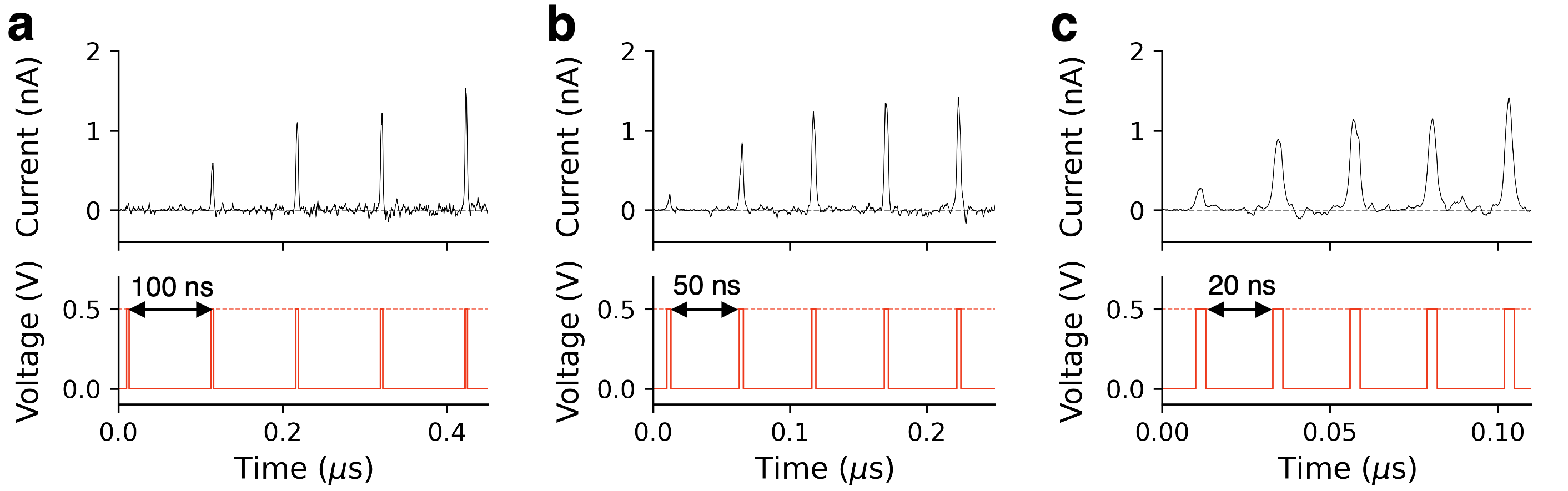}
\caption{\textbf{Ion current potentiation under voltage spikes applied at different intervals.} \textbf{a.} $\Delta t$ = 100 ns \textbf{b.} $\Delta t$ = 50 ns\textbf{c.} $\Delta t$ = 20 ns. The pulse height $V_s$ is set to 0.5 V and $\tau$ = 5 ns.}
\label{SI_fig1}
\vspace{18pt} 
\end{figure*}
Progressive increase (potentiation) of ion current in response to a series of five successive voltage pulses, applied at the intervals of 20 ns, 50 ns, and 100 ns is shown in Fig.~\ref{SI_fig1}. The results show that the hBN membrane with B$_3$N vacancies exhibits consistent current potentiation in all cases.

\vspace{12pt} 
\begin{figure*}[h]
\centering
\includegraphics[width=1.0\textwidth]{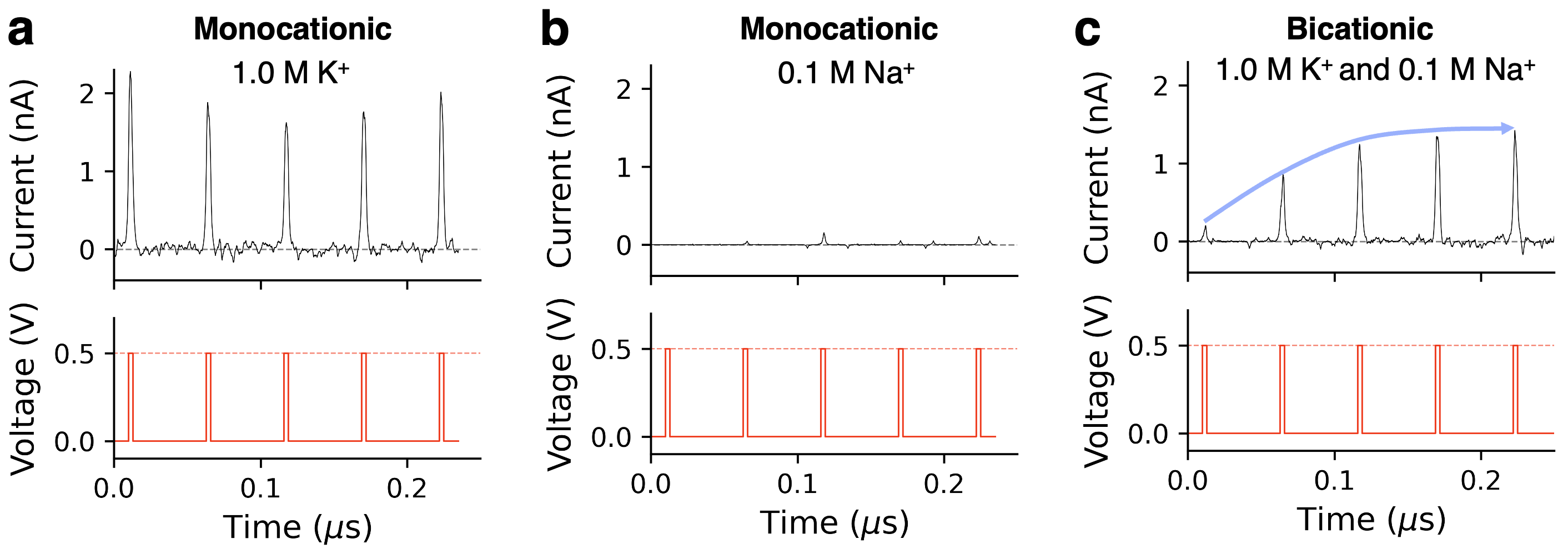}
\caption{\textbf{Monocationic system \textit{vs} bicationic system.} Ion current induced by five voltage pulses for monocationic systems with \textbf{a.} 1.0 M KCl and \textbf{b.} 0.1 M NaCl, and bicationic systems with \textbf{c.} 1.0 M KCl and 0.1 M NaCl. The pulse height $V_s$ is set to 0.5 V and $\tau$ = 5 ns with 50 ns pulse interval.}
\label{SI_fig2}
\vspace{18pt} 
\end{figure*}

Fig.~\ref{SI_fig2} shows the important role of bicationic electrolyte in the synapic-like potentiation. When only K$^+$ ions are present, a uniform response of ion current pulses is exhibited under successive voltage spikes, as shown in Fig.~\ref{SI_fig2}a. In this case, the conductive state of the membrane remains un-modulated, as all pores stay activated in the absence of ions blocking and unblocking pores (Na$^+$, in this case) with significant time delays. Conversely, in the monocationic system featuring only Na$^+$ ions, the simulated ion currents are negligible. It is worth noting that beyond a certain number of pulses, the difference between the analytical asymptotic value of ion conductance and that obtained from individual MD simulations would indeed be a number rooted in the stochastic component of the individual simulations. For an experimental system, however, this stochastic discrepancy is expected to be significantly lower. 

\vspace{12pt} 
\begin{figure*}[h]
\centering
\includegraphics[width=0.75\textwidth]{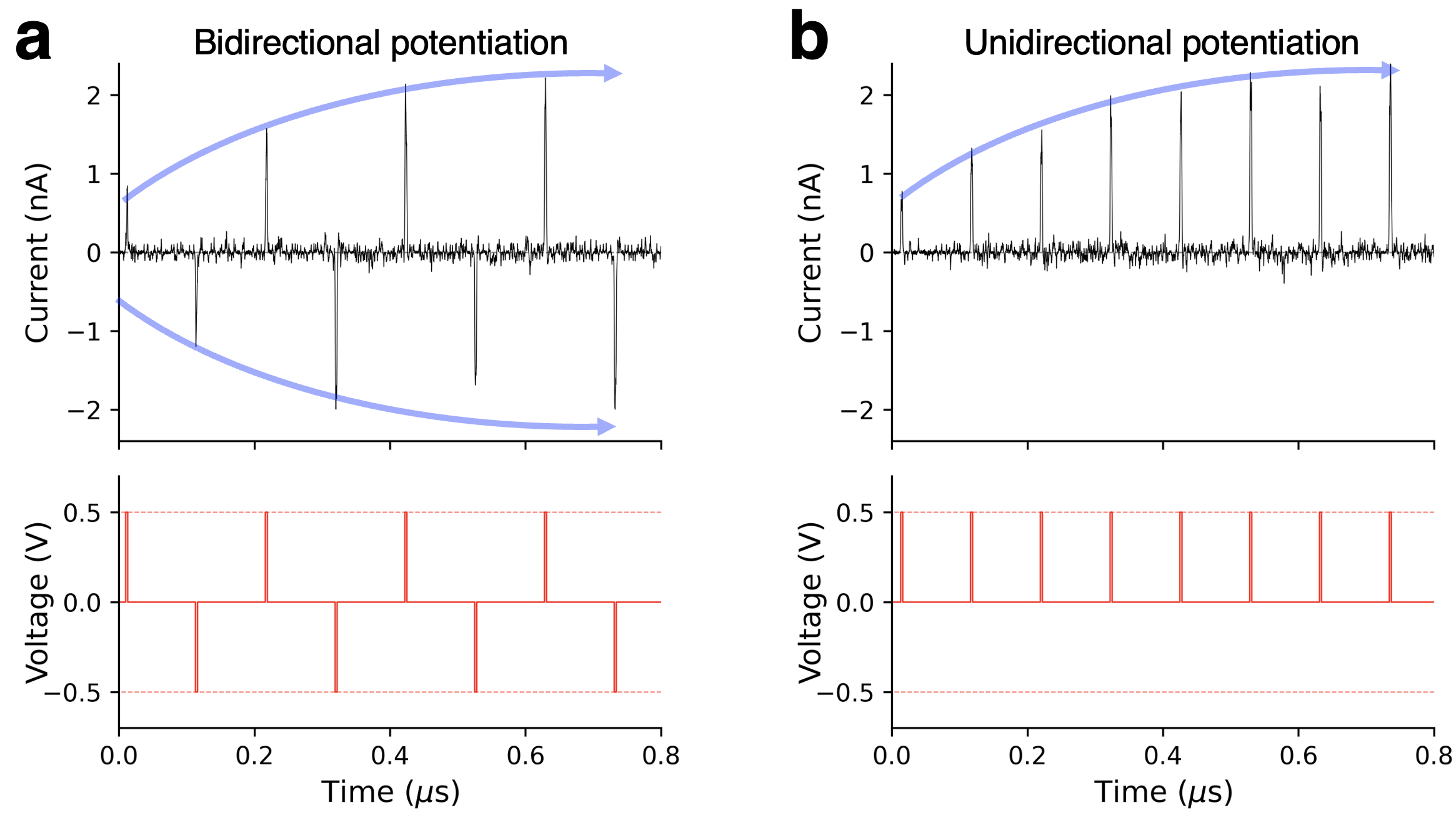}
\caption{\textbf{Bidirectional current potentiation.} A sequence of voltage pulses of \textbf{a.} alternating and \textbf{b.} single polarity. The pulse height $V_s$ is set to 0.5 V and $\tau$ = 3 ns with 50 ns pulse interval.}
\label{SI_fig3}
\vspace{18pt} 
\end{figure*}

As suggested in the discussion presented in the main text, this 2D membrane functions as a bidirectional memristor, as depicted in Fig.~\ref{SI_fig3}. The figure illustrates that the ion current is potentiated regardless of bias voltage polarity, as expected for the presented $Z$-symmetric configuration. 

\vspace{12pt} 
\begin{figure*}[h]
\centering
\includegraphics[width=0.65\textwidth]{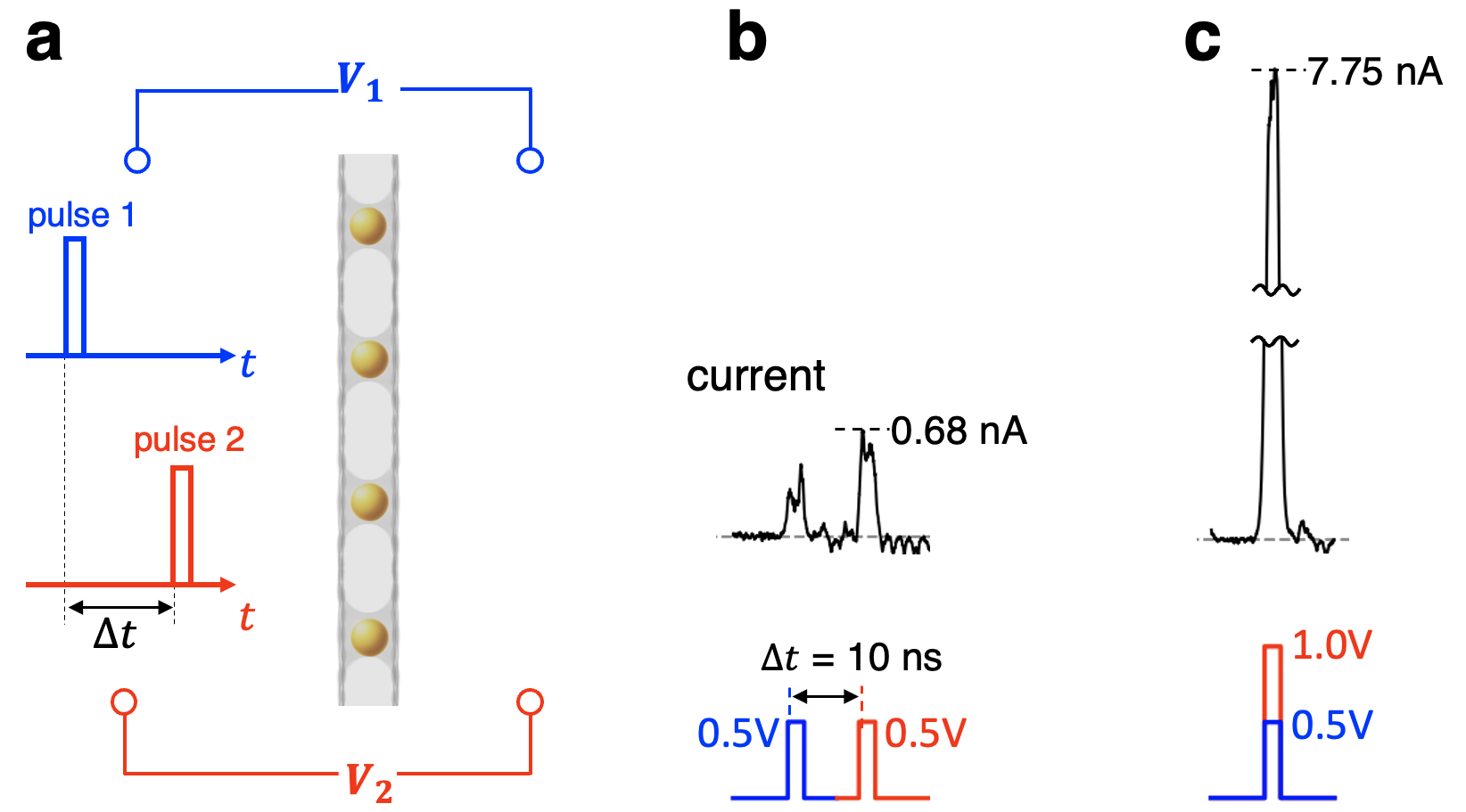}
\caption{\textbf{Spiking time-dependent current.} \textbf{a.} An illustration of two external voltage sources attached in parallel, generating pulses with a timing difference of $\Delta t$. Ion current for the spiking time difference of \textbf{b.} $\Delta t$ = 10 ns and \textbf{c.} $\Delta t$ =0, with a superimposed voltage.
\vspace{24pt} 
\label{SI_fig4}}
\end{figure*}

As described in the main text, both the peak current ($I_{peak}$) and the change in memristive state ($\Delta w$) have a nonlinear dependence on the applied voltage magnitude. Thus, as shown in Fig.~\ref{SI_fig4}, the resulting current peak is significantly higher when two voltage spikes occur simultaneously, compared to a case of two voltage spikes occurring 10 ns apart.

\vspace{12pt}
\begin{figure*}[h!]
\centering
\includegraphics[width=1.0\textwidth]{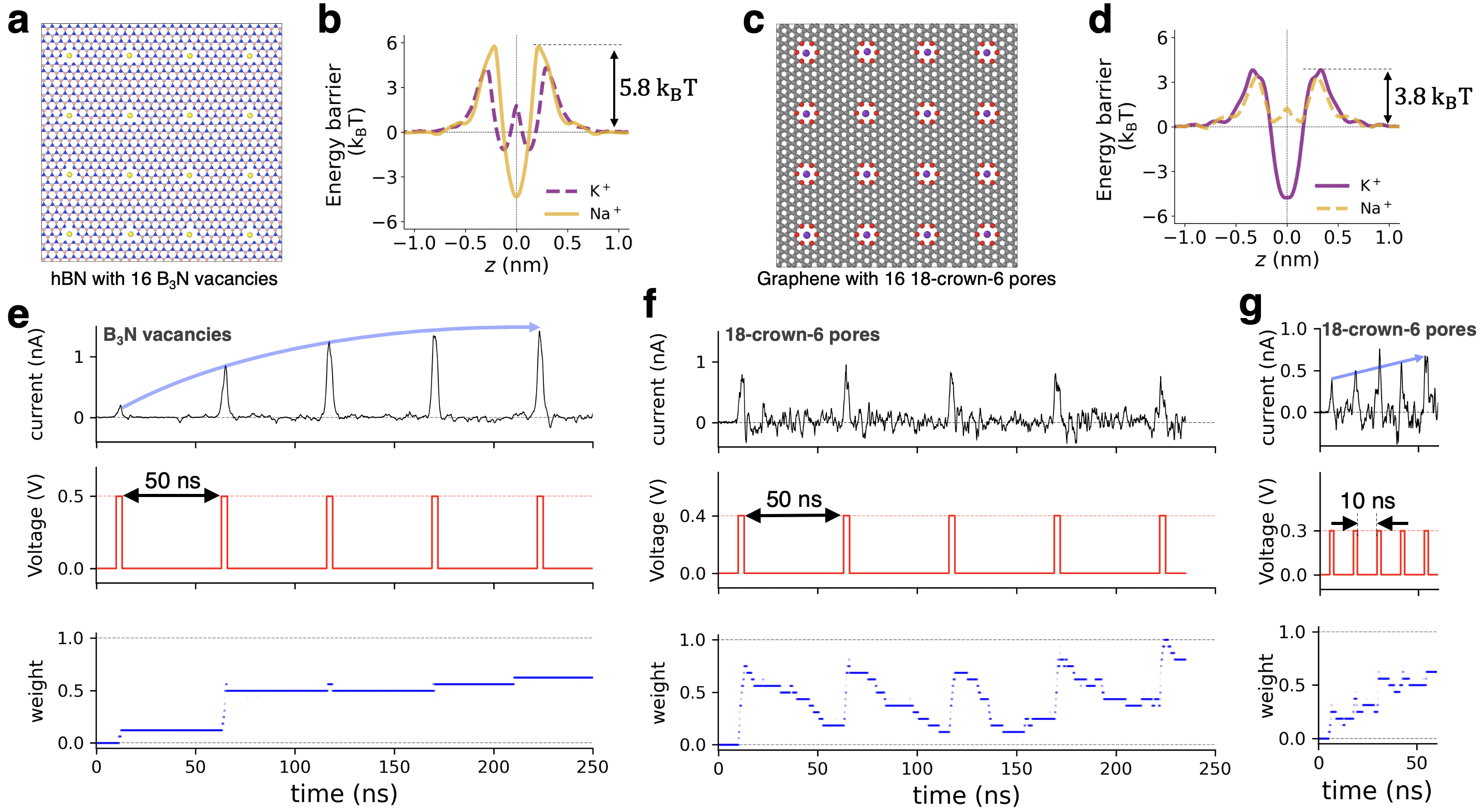}
\caption{ \textbf{Monolayer hBN with 16 B$_3$N vacancies \textit{vs} Monolayer graphene with 16 18-crown-6 pores.} \textbf{a.} A top view of the hBN sheet with 16 B$_3$N vacancies. The colors of the atoms are as follows: Nitrogen - blue; Boron - pink; Sodium ion - yellow. \textbf{b.} Energy barrier along the transport coordinate across B$_3$N vacancy. \textbf{c.} A top view of the graphene sheet with 16 18-crown-6 pores. The colors of the atoms are as follows: Carbon - gray; Oxygen - red; Potassium ion - purple. \textbf{d.} Energy barrier along the transport coordinate across the 18-crown-6 pore. Ion current, voltage, and weight as a function of time for \textbf{e.} the hBN sheet with a 50 ns time interval between pulses, \textbf{f.} the graphene sheet with a time interval of 50 ns, and \textbf{g.} the graphene sheet with a time interval of 10 ns.
\label{SI_fig5}}
\vspace{12pt} 
\end{figure*}

We compared the ion current dynamics exhibited by two distinct 2D nanofluidic memristors when subjected to a series of five consecutive voltage pulses. The first system consists of hBN with an array of B$_3$N defects, as described in the primary discussion of our manuscript and shown in Fig.~\ref{SI_fig5}a. The second system is a single layer of graphene featuring an array of 16 18-crown-6 pores, as shown in Fig.~\ref{SI_fig5}d. This membrane is known to exhibit memristive effects through the same mechanism described in the main text, except with the roles of Na$^+$ and K$^+$ ions are reversed, as suggested by the qualitatively opposite energy landscapes shown in Figs.~\ref{SI_fig5}b, \ref{SI_fig5}e. Specifically, in the case of the 18-crown-6 pore array, Na$^+$ serves as the primary charge carrier across the membrane, while K$^+$ ions act as pore blockers. As observed previously, the hBN membrane with B$3$N vacancies exhibits potentiation in ion permeability under a 50 ns pulse interval (see Fig.~\ref{SI_fig5}c). However, the graphene membrane with 18-crown-6 pores does not show notable potentiation under a 50-ns-long pulse (see Fig.~\ref{SI_fig5}f), due to the considerably faster weight decay between potentiating pulses, as supported by a 2 $k\mathrm{B}T$-lower adsorption energy barrier in the case of crown-like pores in graphene. For a 10 ns pulse interval, the 18-crown-6 pore array exhibits slight ion current potentiation (see Fig.~\ref{SI_fig5}g). This example illustrates different operational pulse frequencies depending on the material, as well as pore composition and geometry of a 2D nanofluidic memristor.
\section*{S2. Adsorption-desorption dynamics of a binary mixture}
The model is similar to that discussed in the main text, except here the competition for the adsorption sites (B$_3$N pores) is included explicitly. For two cation species, the governing system of equations is: 
\begin{equation}
    \begin{cases}
      \frac{dN_1}{dt} = -r_{d,1} N_1 + r_{a,1} (N_{tot} - N_1 - N_2),\\
      \frac{dN_2}{dt} = -r_{d,2} N_2 + r_{a,2} (N_{tot} - N_1 - N_2)\\
    \end{cases}
\label{eq_s1}
\end{equation}

\begin{figure*}[h]
\centering
\includegraphics[width=0.55\textwidth]{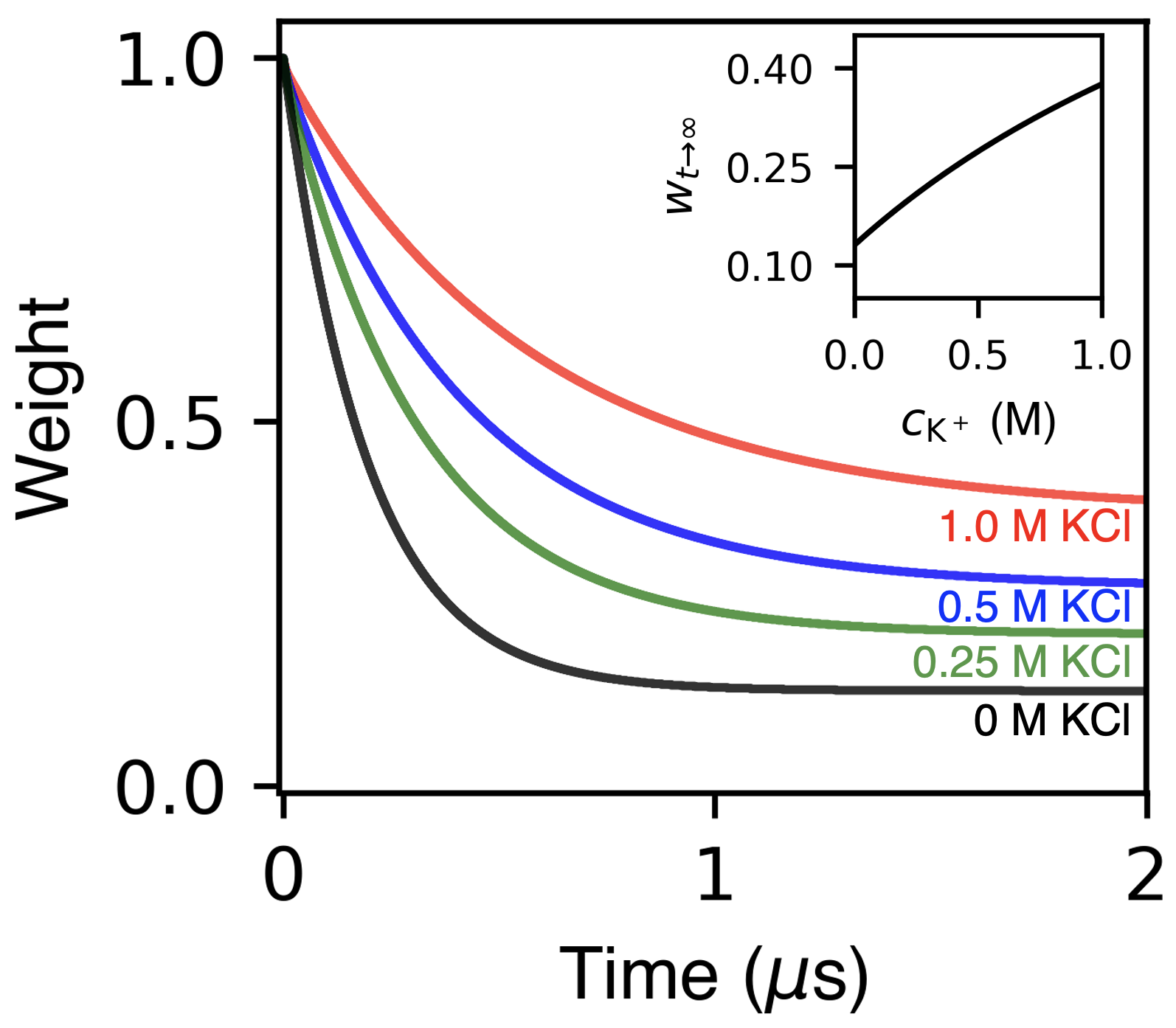}
\caption{\textbf{Weight decaying over time in systems with a binary salt mixture of varying concentrations of KCl and 0.1 M NaCl.} The inset  shows the asymptotic value of $w$ \textit{vs} KCl concentration}
\label{SI_fig6}
\vspace{12pt} 
\end{figure*}

The constants are defined in a simplified way as $r_{d,i} = f_{d,i}\cosh\left({\frac{q\phi}{2k_B T}}\right) \exp\left({-\frac{E_{d,i}}{k_B T}}\right)$ and $r_{a,i} = \kappa_i c_i \exp\left({-\frac{E_{a,i}}{k_B T}}\right)$, where $i=1,2$ refers to Na$^+$ and K$^+$, respectively. For simplicity, we set $f_{d,1} = f_{d,2} = $ 16.5 GHz, $\kappa_1 = \kappa_2 = $ 14.9 GHz$\cdot$M$^{-1}$,  while the barrier heights are taken directly from the data in the main Fig. 1b: $E_{d,1} = 10.1 k_B T$, $E_{a,1} = 5.8 k_B T$, $E_{d,2} = 5.5 k_B T$, $E_{a,2} = 4.3 k_B T$. Unbiased ($\phi = 0$) time-decays starting at $w_0=1$ for several values of KCl concentration and NaCl fixed at 0.1 M are shown in Fig.~\ref{SI_fig6}. Both the decay rate and $w_{t\rightarrow \infty}$ depend on the KCl concentration and can be estimated analytically. We briefly focus on the latter, which is given by $w_{t\rightarrow \infty} = \frac{\lambda_1 (\lambda_1 + \lambda_2)}{\lambda_1 + \lambda_2 + \lambda_1\lambda_2}$, where $\lambda_i = r_{d,i}/r_{a,i}$. This expression is identical to the pore occupancy estimates presented in the supplementary section 3 of our earlier work (26). The corresponding $w_{t\rightarrow \infty}$ as a function of KCl concentration is shown in the inset of Fig.~\ref{SI_fig6}.

\vspace{12pt} 

\clearpage